\documentclass[twocolumn]{aastex7}
\usepackage{multirow}
\usepackage{wasysym}
\usepackage{enumerate}
\usepackage{amssymb} %maths
\defcitealias{Astro2020}{Astro2020}
\shorttitle{Finding Endor with Lunar Eclipses}
\shortauthors{Limbach et al. }

\begin{document}

\title{Exomoons and Exorings with the Habitable Worlds Observatory II: \\Finding `Endor' with Lunar Eclipses}

\correspondingauthor{Mary Anne Limbach}
\email{mlimbach@umich.edu}

\author[0000-0002-9521-9798]{Mary Anne Limbach}
\email{mlimbach@umich.edu}
\affiliation{Department of Astronomy, University of Michigan, Ann Arbor, MI 48109, USA}

%%%%Currently in alphabetical order%%%%%%
\author[0009-0005-7704-5527]{Beck Dacus}
\email{}
\affiliation{Department of Astronomy \& Astrophysics, University of California, San Diego, CA 92093, USA}

\author[0009-0008-5864-9415]{Brooke Kotten}
\email{bkotten@umich.edu}
\affiliation{Department of Astronomy, University of Michigan, Ann Arbor, MI 48109, USA}

\author[0009-0007-5843-6447]{Elizabeth Lane}
\email{}
\affiliation{Department of Astronomy, University of Michigan, Ann Arbor, MI 48109, USA}

\author[0000-0002-0746-1980]{Jacob Lustig-Yaeger}
\email{}
\affiliation{JHU Applied Physics Laboratory, 11100 Johns Hopkins Rd, Laurel, MD 20723, USA}

\author[0000-0003-4816-3469]{Ryan MacDonald}
\email{}
\affiliation{School of Physics and Astronomy, University of St Andrews, North Haugh, St Andrews, KY16 9SS, UK}

\author[0000-0002-3196-414X]{Tyler D. Robinson}
\email{}
\affiliation{Lunar \& Planetary Laboratory, University of Arizona, Tucson, AZ 85721 USA}

\author[0000-0003-2233-4821]{Jean-Baptiste Ruffio}
\email{}
\affiliation{Department of Astronomy \& Astrophysics, University of California, San Diego, CA 92093, USA}

\author[0000-0001-7246-5438]{Andrew Vanderburg}
\email{}
\affiliation{Center for Astrophysics, Harvard \& Smithsonian, 60 Garden Street, Cambridge, MA 02138, USA}

\begin{abstract}
Giant planets in the habitable zone may host exomoons with conditions conducive to life. In this paper we describe a method by which the \textit{Habitable Worlds Observatory} (HWO) could detect such moons: broadband reflected–light lunar eclipses (e.g., { the moon passing into the shadow of the planet}). We find that an Earth–like moon orbiting a Jovian–size planet at 1~au can outshine its host planet near 1~$\mu$m, producing frequent (days time–scale) lunar eclipses with depths of order 50\%. We determine that single eclipse events out to $\sim$12\,pc may be detectable for Earth–like moons around giant planets, down to $0.9R_\oplus$. Detection of smaller moons, $\sim$0.5$R_\oplus$ (corresponding to about the size of Mars or Ganymede), may be possible, but would generally require multiple events for most systems. These several–hour events provide a clear pathway to detecting habitable moons with HWO, given sufficient stare-time on each system to detect lunar eclipses. The occurrence rate of habitable exomoons remains unconstrained, however, making the ultimate yield uncertain. HWO will be capable of placing the first meaningful constraints on the frequency of habitable exomoons around giant planets; if it is non–negligible, HWO could also search for life on these worlds, possibly with lunar eclipse spectroscopy.
\end{abstract}


\section{Introduction}\label{sec:intro}

The Habitable Worlds Observatory (HWO), proposed in the National Academies’ \emph{Pathways to Discovery in Astronomy and Astrophysics for the 2020s} Decadal Report \citep[][henceforth ``\citetalias{Astro2020}'']{Astro2020}, is a concept for a large space telescope operating in the infrared, optical, and ultraviolet. The mission’s primary objective is to locate and characterize habitable worlds outside our Solar System, aiming to directly image at least 25 potentially habitable worlds.

Considerable effort has been focused on estimating how many stellar systems must be observed to detect $\sim$25 habitable zone terrestrial planets \citep{2019arXiv191206219T,2020arXiv200106683G,2024arXiv240212414M}. However, there is currently no literature estimating how many habitable exomoons HWO might discover. Moreover, while a few studies mention the detectability of habitable exomoons with HWO in passing \citep{2014PNAS..111.6871R,2015ApJ...812....5A,2024AJ....168...57L}, HWO's sensitivity to these worlds remains largely unconstrained.

If our Solar System’s rich population of moons is indicative \citep{1989Sci...246.1459B,1999Sci...296...77S,2006RvMG...60..221W,2017JGRE..122..432H,2018RPPh...81f5901D,2023JGRE..12807432C}, many exoplanetary systems may also host diverse satellite systems. Such exomoons, spanning a range of compositions and conditions, could present a broad spectrum of environments, some of which may reside within the habitable zones of their host planets and potentially possess conditions conducive to life \citep{1987AdSpR...7e.125R,2010ApJ...712L.125K,2014AsBio..14..798H,2014OLEB...44..239L,HellerBarnes_2015,2018MNRAS.479.3477H}. While there is uncertainty in $\eta_\oplus$, the frequency of habitable planets \citep{2011ApJ...738..151C,2013ApJ...766...81F,2013ApJ...765..131K,2020AJ....159..248K,2024AsBio..24..916S}, the frequency of habitable exomoons around giant planets is almost entirely unconstrained, as no exomoons have yet been confirmed, and certainly none are known in or near the habitable zone. Even so, occurrence-rate studies indicate that $\sim$5\% of FGKM stars host a giant planet in the habitable zone, and each such planet may harbor multiple moons, implying a potentially large underlying population \citep{2021ApJS..255...14F,2023AJ....165..113R}. To estimate how many habitable moons HWO might detect, we must first determine which potentially habitable-zone moons are detectable with HWO.

In this paper, we explore the detectability of habitable-zone exomoons around giant planets with HWO via lunar eclipses (i.e., the planet passing between the star and the moon). We lay out our method in Section~\ref{sec:methods} and present the resulting detection limits in Section~\ref{sec:Results}. We then discuss some caveats to our findings and the implications of this work in Section~\ref{sec:Discussion}, and conclude in Section~\ref{sec:Conclusion}.

\section{Methods}\label{sec:methods}

Lunar eclipses of exomoons around giant planets should be ubiquitous and frequent \citep{2024AJ....168...57L}. In our Solar System, the three inner Galilean moons of Jupiter (Io, Europa, and Ganymede) are regularly eclipsed by Jupiter’s shadow once every orbital period of the moon, with periods ranging from 1.8\,days (Io) to 7.2\,days \citep[Ganymede,][]{1984AJ.....89..280A,2023MNRAS.526.6145C}. By analogy, exomoons of giant planets in other systems are likely to be close to coplanar with the star–planet orbital plane, reflecting formation in circumplanetary disks and the difficulty of exciting large inclinations for massive, long-lived satellites \citep{2010AJ....140.1168W,2020ApJ...894..143B}.

{ In the case of planets with a highly tilted obliquity, such as Uranus, short-period moons still undergo a lunar–eclipse ``season,'' during which eclipses occur frequently. For example, Miranda, the innermost major moon of Uranus, produces a lunar eclipse every 1.4 days (its orbital period) throughout a $\sim$5.3-year-long eclipse season that recurs every $42$ years (i.e., lunar eclipses are possible for approximately 13\% of Uranus's orbit, when $\sin i < R_{p}/a_{m}$). Therefore, the moons of highly oblique planets remain detectable via lunar eclipses, although the probability of observing the system during an eclipse season is naturally lower than in systems such as Jupiter–Io, where eclipses occur throughout the planet's entire orbit about the Sun. Intriguingly, long-term monitoring of in eclipse-season durations and of changes in lunar eclipse durations could provide strong constraints on planetary obliquities, assuming the moons remain coplanar with the planet’s tilt.}

Detecting moons via lunar eclipses therefore reduces to demonstrating that HWO has sufficient sensitivity to detect these events and monitoring each target for a long enough duration to capture an eclipse, unless planetary variability is significantly larger than observed in our own system, in which case detection may become more challenging.
{ The long monitoring duration required for detection is likely the principal drawback of the method: while short-period moons (e.g., $P \sim 2$\,days) can be detected within a feasible continuous monitoring interval, waiting $\sim$10\,days to capture longer-period eclipses may be impractical. Here, we assume that 2-3\,days is not an unreasonable search timescale, as similar total stare times may be required to detect habitable-zone terrestrial planets in some systems.} However, this science case can often be pursued in parallel with others: long-duration monitoring of gas giants for atmospheric characterization, variability studies, or searches for additional planets (while the giant remains in the field of view) naturally enables exomoon eclipse searches at minimal additional cost.
 Furthermore, each of the Solar System’s giant planets hosts at least one moon with a relatively short ($<3$\,day) orbital period, so we proceed with the understanding that the yield of this technique will be limited by the feasible continuous monitoring duration per system.

To compute lunar eclipse depths, we require spectral models for both the moon and its host planet. In this work we adopt modern Earth as an exomoon analog and a Jupiter-mass planet at 1\,au as the host. We recognize that this Earth-like assumption is simple. Models of Earth through time and considerations of “life as we don’t know it” \citep{2018haex.bookE.189O,2018SciA....4.5747K,2018haex.bookE..70P,2024AsBio..24S.186G} suggest habitable worlds (both planets and moons) could host biospheres with spectra unlike present-day Earth. Nevertheless, for the purposes of this study we proceed with the Earth-like moon orbiting a Jupiter-mass planet at 1\,au as a clear baseline case.

For the host planet’s reflected–light SED we used the \textit{Reflection Spectra Repository for Cool Giant Planets} \citep{2018ApJ...858...69M,reflection_spectra_cool_giant_planets_2019}, specifically the HDF5 file \texttt{Cool\_giant\_albedo\_database.hdf5}. The database contains 65520 model reflection spectra spanning metallicity (i.e., $1$–$100\times$ solar), gravity ($g=1$–$100~\mathrm{m\,s^{-2}}$), effective temperature ($T_{\rm eff}=150$–$400$~K, 10~K steps), and cloud sedimentation ($f_{\rm sed}=1$–$10$). The grid was computed with \texttt{PICASO} reflected–light albedo code \citep[including Raman scattering,][]{2019ApJ...878...70B}. For our fiducial Jupiter analog we extracted the model with $\log(m)=0.5$, $\log g=3.4$ ($g\approx25~\mathrm{m\,s^{-2}}$, same as Jupiter), $T_{\rm eff}=300$~K, and $f_{\rm sed}=10$, and used the geometric–albedo output provided in the HDF5 file.

\begin{figure}
  \begin{center}
    \includegraphics[width=0.45\textwidth]{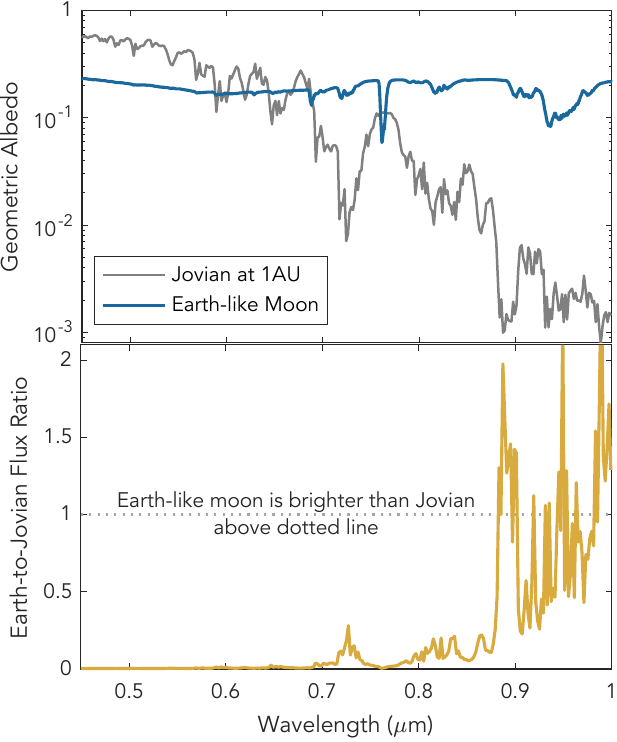}
  \end{center}
  \caption{{\bf Top:} Modeled geometric albedo of a warm Jupiter (300\,K) at 1\,au around a Sun-like star \citep{2018ApJ...858...69M,reflection_spectra_cool_giant_planets_2019} and geometric albedo of Earth \citep{2011AsBio..11..393R}. {\bf Bottom:} A Earth-like moon outshines the warm Jovian planet in regions of the spectrum beyond 0.85\,$\mu$m { because the planet’s low albedo at these wavelengths is set by strong methane and water absorption bands}.}
  \label{EarthAndJup}
\end{figure}

For the moon analogue, we used disk–integrated Earth spectra from the Virtual Planetary Laboratory (VPL\footnote{\url{https://live-vpl-test.pantheonsite.io/models/vpl-spectral-explorer/}}) spectral modern Earth model \citep{2011AsBio..11..393R}, validated against NASA’s EPOXI Earth observations. We obtained the quadrature-phase spectrum from the VPL spectral database \citep{2011AsBio..11..393R} and converted the provided disk-averaged reflectance at quadrature to geometric albedo so that the units match those of the planet model (geometric albedo rather than disk-integrated albedo at quadrature). Later in the manuscript, when computing flux ratio at a given phase, we convert both the planet and moon models from geometric albedo to disk-integrated reflectance using the appropriate phase functions for each body.

The geometric albedo as a function of wavelength is shown in the top panel of Figure~\ref{EarthAndJup}. The bottom panel plots the relative reflected flux, computed as the Earth-like moon spectrum divided by the 300~K Jovian spectrum,including the cross-sectional area factor \((R_{\rm m}/R_{\rm p})^{2}\). The moon is notably brighter than the planet ($>1.5\times$) in several windows from 0.85-1.0~$\mu$m, where strong H$_2$O and CH$_4$ absorption depresses the giant planet continuum. However, we see from \cite{2015ApJ...812....5A}, which has a model out to 3$\mu m$ (see their figure 5), there are several additional spectral regions beyond 1\,$\mu m$ where the Earth-like moon outshines the giant planet. If our giant–planet models are correct, this realization is quite intriguing: an Earth–like moon can outshine a giant planet at 1\,au in the near–infrared. This has several intriguing implications, but for this study we return our focus to lunar eclipses.
Consequently, a total lunar eclipse removes the dominant source of reflected light in this hypothetical blended planet-moon system spectrum, producing a pronounced drop in flux. %We note that there currently are not publicly available reflected light models of warm Jupiters beyond 1.00~$\mu$m and hence our spectrum does not extend to longer wavelengths.

\begin{figure*}
  \begin{center}
    \includegraphics[width=0.92\textwidth]{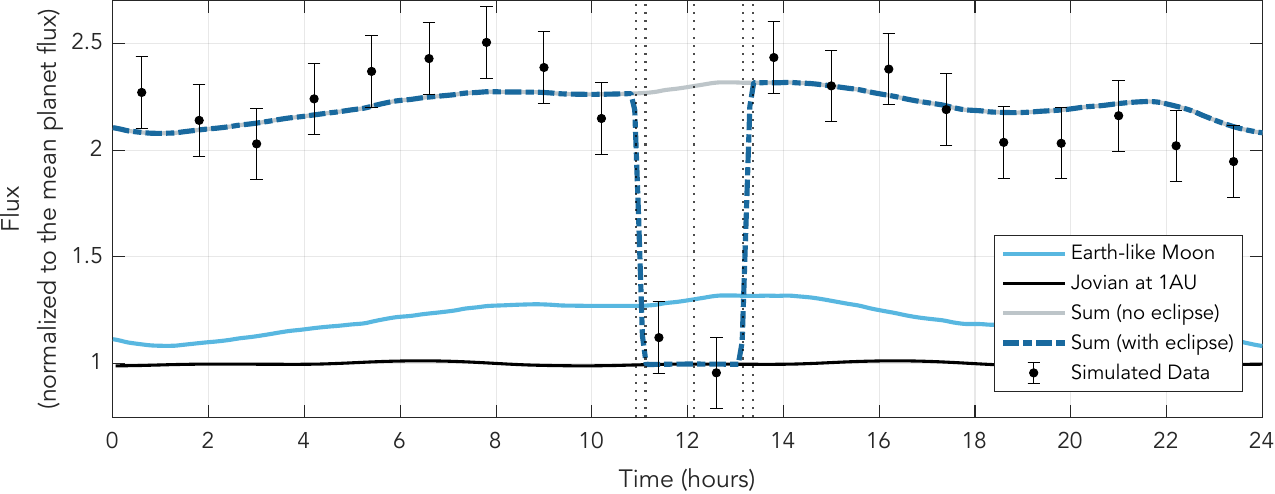}
  \end{center}
  \caption{Lunar–eclipse model for an Earth–like moon (light blue) and a Jupiter–sized planet (black), where the Earth–like moon is $\sim$1.2$\times$ brighter (on average) than the giant planet. Blended time–series photometry with (dark–blue dash–dotted) and without (light gray) the lunar eclipse is shown. Both light curves include measured rotational variability for Earth and Jupiter in the near IR. The deep eclipse near 12~h corresponds to a $>50\%$ drop in the flux (i.e., a $>10^{-10}$ decrement). In this example the Earth–like moon follows an Io–like orbit (1.8~d, edge-on), yielding a few–hour eclipse duration at 1.8~d cadence. Notably, the moon’s variability dominates over the planet’s. The simulated data (black error bars) assumes the eclipse is detected with an SNR = 9.1 as discussed in section \ref{sec:Results} for a Jovian hosting a earth-like moon at 5\,pc. The simulated points have a cadence of 1.2\,hr, corresponding to an SNR of 7.8.}
  \label{AnEndorEclipse}
\end{figure*}

{ Figure~\ref{EarthAndJup} shows that the lunar eclipse depth of an Earth-like exomoon orbiting a giant planet at 1\,au should exceed 50\% of the combined planet+moon reflected flux in the bandpass near $\lambda = 1.00~\mu$m, corresponding to the removal of the moon's contribution when it is brighter than the planet.} Next, we illustrate this by building a light curve. We include variability from both the planet and the moon to demonstrate that the lunar eclipse dwarfs the intrinsic variability of either body. We again leverage EPOXI Earth data, specifically “Observation 1” timeseries data from \cite{2011AsBio..11..907L} at 850~nm, the closest available light curve wavelength to our region of interest near 1.00~$\mu$m. This light curve is the light–blue line in Figure~\ref{AnEndorEclipse}. 

In this example, the Earth-like moon is $\sim$1.2$\times$ brighter than its Jovian host. A moon-to-planet flux ratio of $\approx$1.2 is perhaps even a conservative choice in the near-infrared: models exceed this level in several very narrow bands shortward of $1~\mu$m and across broad ($\sim20\%$) bandpasses longward of $1~\mu$m \citep{2015ApJ...812....5A}. In the spectrum shown in Figure~\ref{EarthAndJup}, the moon-to-planet brightness ratio is $\approx 1.2$ in the $0.98$–$1.00~\mu$m bandpass; from the NIR spectrum of \cite{2015ApJ...812....5A}, we estimate that the ratio remains roughly the same in the wider $0.98$–$1.18~\mu$m ($\sim$20\%) bandpass.

We leverage HST/WFC3 light curves of Jupiter at 900~nm \citep{2019AJ....157...89G}, stitching together three consecutive rotations (Jupiter’s rotation period is $\sim$10~h). The planetary light curve is shown as the black line in Figure~\ref{AnEndorEclipse} and is normalized to have a mean of one (note that the moon's flux was then normalized relative to the planet's). The planet is notably less variable than the moon. We note, however, that these variability amplitudes are drawn from a significantly colder planet: Jupiter’s effective temperature is $\sim$125~K \citep{2023RemS...15.1811R}, whereas our model planet is 300~K, so variability patterns on a warmer giant planet could be different. For context, WISE~0855, a free-floating object a few times more massive than Jupiter with $T_{\rm eff}\sim300$~K, exhibits $\sim$5\% variability in the near-IR \citep[][a bit larger than what we see from Jupiter]{2016ApJ...832...58E}, although it cannot be observed in reflected light since it has no host star.

The dark–blue dash–dotted curve in Figure~\ref{AnEndorEclipse} shows the combined planet+moon light curve with a lunar secondary eclipse injected near the 12~h mark. In this example we adopt Io’s orbital period (1.8~d) and assume an edge–on event, which yields a total eclipse duration of $\simeq$2.2~h. The system flux drops by $>$50\% during eclipse, consistent with the broadband moon/planet flux ratio of $\sim$1.2 (depth $=r/(1+r)\!\approx\!0.55$, where $r$ is the ratio of moon-to-planet flux). The planet and moon rotational variabilities are included but are much smaller than the eclipse signal. We note that only the lunar eclipse (the moon passing into the shadow of the planet) not the stellar eclipse (the moon passing between the star and planet) is shown in Figure~\ref{AnEndorEclipse}. Given the 1.8~d orbital period, the stellar eclipse would occur near 34~h (outside the plotted window) and would be undetectable in this bandpass because its quadrature depth is only $\sim 1\%$ (far smaller than the $>50\%$ depth of the lunar eclipse) though it may be detectable in other spectral bands, which we do not investigate here.

\section{Results}\label{sec:Results}

\begin{figure*}
  \begin{center}
    \includegraphics[width=0.92\textwidth]{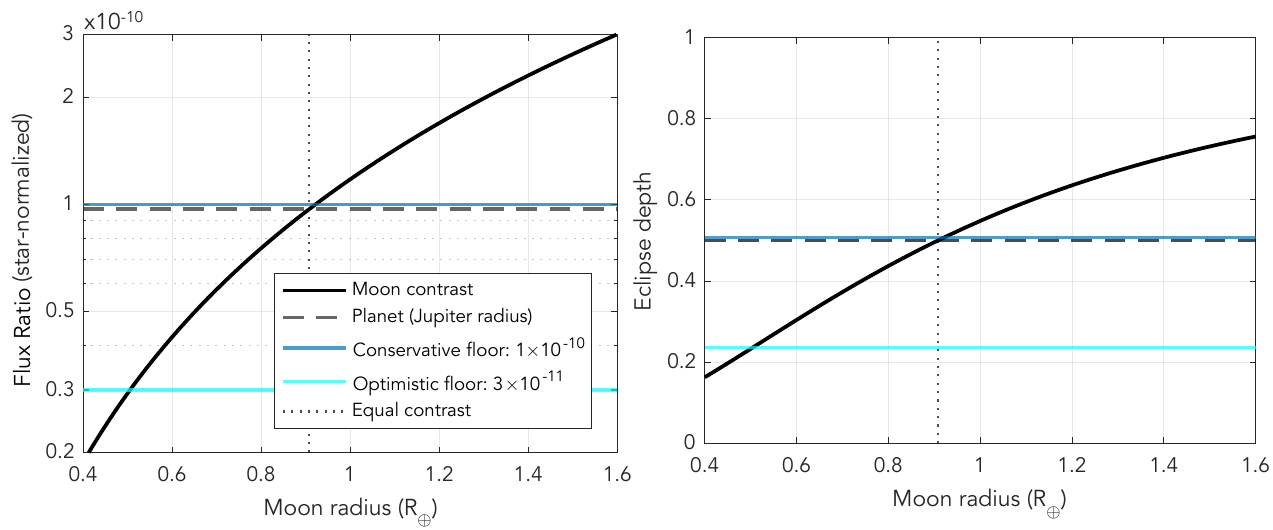}
  \end{center}
  \caption{\textbf{Left:} Moon–to–star flux ratio (black solid line) as a function of moon radius, compared to the planet–to–star flux ratio (black dashed horizontal line; Jupiter–size planet). Assumes a Sun-like star and a 1~au orbit for the planet-moon about the star. Where the moon is brighter than the planet ($R_{\rm moon}\gtrsim0.9\,R_\oplus$ and flux ratio $\gtrsim10^{-10}$; dark–blue and black dotted lines), lunar–eclipse detection will be possible with a single eclipse out to $\sim$12\,pc. Smaller moons (down to $0.5\,R_\oplus$) may be detectable depending on HWO’s post–processed contrast floor \citep[light–blue line at $3\times10^{-11}$,][]{2024ApJS..272...30H}, but will likely require multiple eclipse measurements. \textbf{Right:} Lunar eclipse depth (black line) versus exomoon radius for a Jupiter–sized host.}
  \label{ContrastsAndDepths}
\end{figure*}

We have shown that near 1~$\mu$m an Earth-like moon can outshine its host giant, and the resulting lunar eclipse depth is a strong signal, exceeding 50\% over a few hours during the event (Figure~\ref{EarthAndJup}). Based on this, we now ask: which lunar eclipses would HWO be able to detect? Because the moon dominates the flux in this bandpass, if HWO can detect the moon in isolation within the observing time spanning one eclipse in the bandpass of interest, it should also detect the eclipse in a moon–planet system as the planet would contribute comparatively little noise. To examine this more carefully, we plot the flux ratio (relative to the star) of the planet and of a moon as a function of moon radius. Here we convert both bodies to disk-averaged reflectance at quadrature before computing their flux ratios relative to the star. The resulting flux ratios as a function of exomoon size (the planet as a horizontal dashed black line and the moon as a solid black line) are shown in the left panel of Figure~\ref{ContrastsAndDepths}. A dotted vertical line at $\simeq0.9\,R_\oplus$ marks where the moon and planet have equal brightness. For smaller moons the moon is fainter than the planet; the corresponding eclipse depths are shown in the right panel, where the depth drops below 50\% at the same $\sim0.9\,R_\oplus$ equality point.

\begin{figure}[b]
  \begin{center}
    \includegraphics[width=0.41\textwidth]{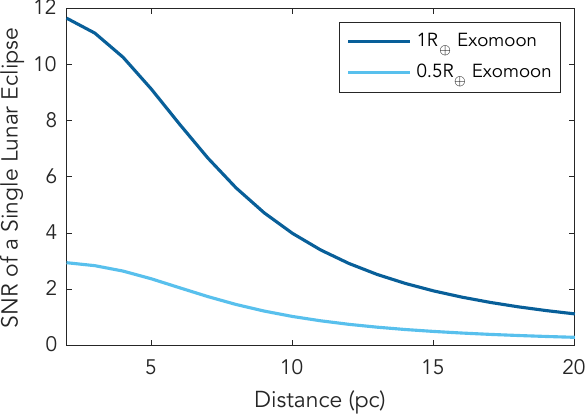}
  \end{center}
  \vspace{-3mm}
  \caption{Signal-to-noise ratio (SNR) for a single lunar eclipse of an Earth-sized exomoon (dark blue) and a $0.5\,R_\oplus$ exomoon (light blue). Earth-sized moons remain detectable out to $\sim$12\,pc in a single event, whereas smaller moons generally require multiple eclipses to achieve detection.}
  \label{SNRplot}
\end{figure}

For moons that produce a $10^{-10}$ change in flux ratio (i.e., a $\sim0.9\,R_\oplus$ moon) over the few hour eclipse, we can estimate the signal-to-noise of the detection of such an event assuming a moon-to-planet flux ratio of 1.2 in the 0.98-1.18$\mu m$ spectral region. To estimate the SNR we might expect for a 2.2 hour transit, we modified {\tt MinimalCoron.py} in the {\tt hwo-tools}\footnote{\url{https://github.com/spacetelescope/hwo-tools}} python package with the following parameters: $R = 5$ (20\% bandpass), $d = $5, 10, 15~pc, integration time = 2.2\,hr and used a 8\,m telescope. We adjusted the stellar brightness, rather than adopting the V-band value used in {\tt MinimalCoron.py}, with the same tool\footnote{\url{https://irsa.ipac.caltech.edu/data/SPITZER/docs/dataanalysistools/tools/pet/magtojy/}}, and derived the J-band flux assuming an absolute solar magnitude of $J=3.65,\mathrm{mag}$ \citep{2018ApJS..236...47W}. All other simulation parameters were consistent with those in {\tt MinimalCoron.py}. We note that this routine does not include detector noise. With this calculation, we found that detecting a $1.2\times10^{-10}$ flux ratio change (what we anticipate for an earth-sized object from figure \ref{ContrastsAndDepths}) in the 2.2\,hr eclipse duration was detectable with a signal-to-noise ratio of 9.1, 4.0 and 1.9 for $d = $5, 10, and 15~pc, respectively. The expected SNR on an earth-sized moon from $d = $2-20\,pc is shown in Figure \ref{SNRplot}.

{ We conclude that it should be possible to detect earth-sized moons via lunar eclipses with HWO.} In this regime the moon’s flux ratio relative to the host star is of order $10^{-10}$ (dark-blue line in both panels). Below that size the moon’s flux ratio falls beneath $10^{-10}$ and, coincidentally, the planet becomes brighter. It is possible that moons remain detectable down to $\sim3\times10^{-11}$ with post-processing as this will likely be around the limiting detectable flux ratio with HWO \citep{2024ApJS..272...30H}, but this is approximate and will depend on HWO’s final design parameters. Assuming the photon noise from the host planet is negligible, a single eclipse of a $0.5\,R_\oplus$ ($3\times10^{-11}$ flux ratio) moon would be detected at a signal-to-noise of 2.3 at 5\,pc. Therefore, such small moons will likely require multiple eclipse events for detection, especially at larger distances. The lunar eclipse SNR for a $0.5\,R_\oplus$ moon from 2-20\,pc is also shown in Figure \ref{SNRplot}. In the most optimistic case, moons as small as ${\sim}0.5\,R_\oplus$ (about the size of Mars or Ganymede) could be detectable. However, we note, habitability of smaller worlds may be unlikely on theoretical grounds \citep{ward_brownlee_rare_earth_2000,2007AsBio...7...66R,2014PNAS..11112628M,2019ApJ...881...60A,2021PNAS..11801155T}. In the case of Mars specifically, although it is thought to have once been potentially habitable for a short time, with evidence suggesting liquid water on the surface and active volcanism, its small size led to the rapid cooling and solidification of its core \citep{2001Natur.412..214S,2007SSRv..129..279D}. This in turn shut down tectonic activity and caused the loss of its global magnetic field. Without this protective magnetic shield, the solar wind gradually stripped away Mars’s atmosphere. As the atmosphere thinned, surface pressure and temperatures declined, rendering the planet cold, arid, and inhospitable to sustained biological development, an outcome that suggests worlds smaller than ${\sim}0.5,R\oplus$ may not retain habitable conditions long enough for complex life to emerge \citep{2013AsBio..13..887W}.

We adopt a detection range of ${\sim}0.9\,R_\oplus$ (conservative floor) to ${\sim}0.5\,R_\oplus$ (optimistic floor), at a wide range of orbital separations given sufficient stare time to catch a lunar eclipse. In principle (e.g., with very long stare times), this encompasses nearly all earth-like exomoons that exist around giant planets in the habitable zone, implying that with enough observing time to capture lunar eclipses, HWO is sensitive to habitable exomoons via the lunar-eclipse method near 1~$\mu$m.

\section{Discussion}\label{sec:Discussion}

\begin{figure}
  \begin{center}
    \includegraphics[width=0.45\textwidth]{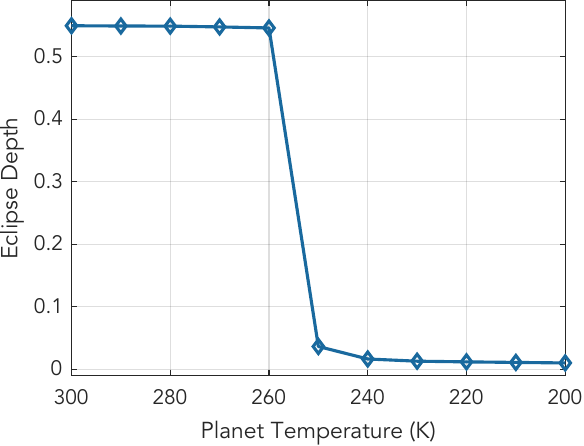}
  \end{center}
  \caption{Eclipse depth (y-axis) at 1 micron for an earth-sized moon around a Jovian planet as a function of temperature of the Jovian planet (x-axis). There is a steep drop near the outer edge of the habitable zone, when a giant planet reaches an effective temperature of T$\sim$250\,K water clouds start forming in the upper atmosphere and raise the albedo causing the eclipse depth to be much less favorable (and likely undetectable) for HWO. Thus, lunar eclipses may only an effective exomoon detection technique for moons that orbit giant planets warmer than 250\,K.}
  \label{TempMatters}
\end{figure}

We have used models of giant planets at 1\,au, not direct measurements, so the true reflected–light albedos of warm giants remain uncertain until verified with observations. Within the model set used here, we find that the low planet albedo near 1~$\mu$m persists across the cloud covers, gravities, and metallicities we tested; however, it weakens significantly and abruptly at lower effective temperatures (near the outer edge of the habitable zone, $T_{\rm eff}\approx250$~K). This trend is shown in Figure~\ref{TempMatters}, which plots eclipse depth versus planet temperature. The sharp drop at $\sim$250~K in the models is likely more subtle in reality { (the water-cloud opacity is model-dependent and the real break is smoother)}, but the implication is the same: once the giant planet is this cool (or cooler), this 1~$\mu$m wavelength range is no longer a favorable place to search for moons via lunar eclipses. This will impede the ability of this technique to detect moons on the outskirts of the habitable zone. On the warmer end (up to the 400~K limit of this model set), the large lunar–eclipse depths persist, suggesting that lunar–eclipse searches should work to the inner edge, and likely even inward, of the habitable zone \citep[though large exomoon occurrence rates are demonstrated to be low at very close star-planet separations,][]{2018AJ....155...36T}.

We have a handle on $\eta_\oplus$, but we have not yet detected any exomoons, so we cannot even begin to extrapolate to an frequency of habitable exomoons around giant planets. HWO will likely constrain this. In our Solar System we see moon-to-planet mass ratios up to $\sim10^{-4}$ \citep[e.g., Titan/Saturn,][]{2006Natur.441..834C} around giant planets. If such large moons are common, we would expect Mars-sized moons around Jupiter-mass planets \citep[potentially large enough to support life,][]{2019ApJ...881...60A} and Earth-sized moons around planets near the brown–dwarf/planet boundary (e.g., $\sim$13~$M_{\rm Jup}$ planets), which are certainly large enough to be habitable.

{ Of the 164 Tier~1 target stars identified for HWO, 144 lie within 20~pc \citep{2021ApJS..255...14F}, the distance to which we have demonstrated exomoon detection is possible \citep[][]{2025PASP..137j4402T}. Based on the habitable zone giant planet ($30$--$6000\,M_{\oplus}$) occurrence rate of $\sim$10\%, we would expect roughly 14 giant planets in the habitable zones of these HWO target stars. The number of detectable moons per planet, however, is highly uncertain. Both theoretical studies \citep[e.g.,][]{2018MNRAS.480.4355C,2021MNRAS.504.5455C,2020MNRAS.499.1023I} and our Solar System suggest that multiple moons per giant planet are typical, though only a small subset are likely to be sufficiently large to detect with this proposed technique. An optimistic upper limit, assuming that every habitable-zone giant planet hosted one detectable and characterizable exomoon, would imply $\sim$14 such systems, a significant fraction of the total population of potentially habitable worlds that HWO is designed to identify and study (e.g., at least 25 habitable zone exoplanets). More realistically, only a minority of these habitable zone giant planets are likely to host moons large enough for robust detection and characterization. Nevertheless, even if only 2--5 habitable-zone giant planets host such satellites, these exomoons could constitute a non-negligible fraction (potentially $\sim$10--20\%) of all habitable worlds accessible to HWO. Moreover, the habitable zone for moons may extend beyond the classical habitable zone for planets due to additional heat sources such as tidal dissipation \citep{2013AsBio..13...18H}. Including giant planets on wider orbits in a potential search for habitable moons could therefore increase the number of viable host systems by several times beyond those expected within the stellar habitable zones alone.}

Exomoons are a place where we should think “outside the box" about what HWO can find. Practically, that argues for (1) keeping stars with habitable-zone giant planets on the target list; (2) planning how to conduct a search for habitable exomoons; and (3) determining how we will characterize any candidates once found. Lunar eclipses may not be the only—or even the best—approach for searching. They are very sensitive but time-inefficient for a blind search. However, dedicated monitoring of large giant planets for lunar eclipses is likely to be scientifically productive, particularly if HWO is sensitive to moons as small as \(0.5\,R_\oplus\). Moreover, such monitoring can be paired with other science cases, e.g., rotational/variability studies of the giant planet, or when the giant remains in the field of view during long integrations targeting smaller planets, so that exomoon searches proceed at minimal additional cost.
 Given that most large giant planets may host such moons \citep{2021MNRAS.504.5455C}, it is quite plausible that this extended stare time will frequently result in exomoon detections. In short, with thoughtful strategy and sufficient time on target, HWO can open a new window on habitable worlds: not just planets, but moons.

\section{Conclusion}\label{sec:Conclusion}

In this paper we explored the detectability of Earth-like exomoons around giant planets with HWO using the lunar–eclipse detection method. We find that:
\begin{itemize}
    \item Reflected-light warm giant planet models combined with the known measured albedo of earth suggest that earth-like moons will outshine giant planets (that are warmer than T$\sim$250\,K) in the near infrared.
    \item HWO will likely be capable of detecting Earth-like moons in the habitable zone via lunar eclipses down to moons the size of $\sim$0.9$\,R_\oplus$ (with a single eclipse) and possibly $\sim$0.5$\,R_\oplus$ (with multiple lunar eclipse events), given sufficient stare time to catch an eclipse(s) (e.g., days).
    \item HWO can use this method (and complementary approaches) to provide the first meaningful constraints on the frequency of habitable exomoons around giant planets { via a targeted reconnaissance observing campaign of approximately 14 giant planets in the habitable zones of HWO target stars.}
    \item Substantial additional work is needed to optimize HWO habitable zone exomoon search strategies and characterization. This is an exciting opportunity to search for life as we don’t know it with HWO and to expand the parameter space in which it can pursue its primary mission: searching for life on habitable worlds beyond our Solar System.
\end{itemize}

\section*{Acknowledgment}
We thank the anonymous referee for their constructive comments, which significantly improved the clarity and quality of this manuscript. MAL also thanks Nell Greenfieldboyce for reminding her how deeply the public cares about this topic, which provided the inspiration and motivation to complete this work. This research has made use of the NASA Exoplanet Archive, which is operated by the California Institute of Technology, under contract with the National Aeronautics and Space Administration under the Exoplanet Exploration Program.

\facilities{Habitable Worlds Observatory}.

\software{{\tt hwo-tools} \citep{hwo-tools}, {\tt astro.py} \citep{2013A&A...558A..33A, 2018AJ....156..123A, 2022ApJ...935..167A}, {\tt numpy.py} \citep{2020NumPy-Array}, {\tt MATLAB}}

\bibliographystyle{aasjournalv7}
\bibliography{main}{}

\end{document}